\documentstyle[preprint,prl,aps]{revtex}

\begin{document}

\draft

\title{Non-Fermi Liquid Behaviour in $d = 2$ Spatial Dimension
}
\author{A. Ferraz$^1$, S. L. Garavelli$^{1,2}$ and T. Saikawa$^1$}
\address{
$^1$International Centre of Condensed Matter Physics\\
Universidade de Brasilia - 70919-970 \\
Brasilia - DF  Brasil}
\address{
$^2$ Departamento de Fisica e Quimica\\
Universidade Catolica de Brasilia\\
Brasilia - DF  Brasil}

\date{\today}
\maketitle

\begin{abstract}
We consider a local effective model for fermionic low lying excitations in a
metal. Introducing a boson auxiliary field and taking into account that the
most significant interactions between quasiparticles arise for those which
are near a given point on the Fermi surface, we find that the effective
boson-fermion interaction is singular for a spatial dimension $d=2$. We
regularize the theory and apply the renormalization group method. As we vary
the coupling constant as a function of an energy scale parameter we show
that the model undergoes a non-perturbative regime in which the Fermi liquid
theory is no longer valid.
\end{abstract}

The normal phase of high-$T_c$ superconductors and some heavy fermions
materials, at low temperature, have in common the fact that some of their
electronic properties cannot be explained by Landau Fermi liquid theory \cite
{Batl,Varma,Anderson,Pines}. The unusual metallic properties, for a spatial
dimension $d=1$, is described well by the so called Luttinger liquid which
possesses a Fermi surface ($S_F$) but no conventional quasiparticles
low-lying excitations. However, for larger spatial dimensions, using
renormalizations group (RG) arguments, several workers have shown that for
any hamiltonian with non-singular interaction the Fermi liquid phase is
always stable and the theory is asymptotically free at the Fermi surface 
\cite{Shankar,Metzner} . As well known a Fermi liquid requires a sharp Fermi
surface characterized by a finite discontinuity in the momentum distribution
function, and one of its features is the logarithmic singularity which
appears in the scattering amplitude for a pair of quasiparticles in the
Cooper channel. This is the only channel in which interactions are not
suppressed by powers of a scale parameter when the physical system
approaches the $S_F$ . As a result for a Fermi liquid all quasiparticle
interactions are `irrelevant' unless the pair of interacting particles has a
total momentum strictly zero. In other words the significant interactions
take place for the fermionic excitations defined around a point on $S_F$ are
produced by the other excitations which are located this same point on the
Fermi surface \cite{Wilczek}. Taking this into account in this work we
consider an effective lagrangian for localized quasiparticles around a given
point on $S_F$.
Using an auxiliary boson field introduced earlier\cite{Ferraz} we show
that for fermionic excitation located around a given point on the Fermi
surface, the effective boson-fermion interaction is singular at $d=2$.

Using renormalization theory we introduce bare fields, chemical potential
and coupling constant and regularize the model to all orders in perturbation
theory. Next, introducing an energy scale parameter, we establish the
renormalization group equation for the renormalized fermion-boson coupling.
Finally integrating the RG equation we show that the physical system
undergoes a non-perturbative regime in which the renormalized coupling grows
indefinitely as the physical system approaches this single point on the
Fermi surface. This renders Fermi liquid theory unapplicable since there is
no longer a trivial infrared fixed point \cite{Benf}.

Let us begin by considering quasiparticles in $d=2$ as described by the
fermionic Lagrangian

\begin{equation}
\label{eq1}{\cal L}=\sum_\sigma \Psi _\sigma ^{+}\left( i\partial _t\text{ }+
\text{ }\frac 1{2m^{*}}\nabla ^2\text{ }+\text{ }\mu _\sigma \right) \Psi
_\sigma \text{ }-\frac U2\Psi _\sigma ^{+}\Psi _{-\sigma }^{+}\Psi _{-\sigma
}\Psi _\sigma 
\end{equation}
where $\mu _\sigma $ is the chemical potential for the spin $\sigma $
quasiparticle, $m^{*}$ is the quasiparticle effective mass and $U>0$ is the
four-fermion coupling constant. The quartic fermionic interaction term is
cumbersome to work with in perturbation theory. For convenience we introduce
a real bosonic auxiliary field $\phi $ instead in such a way that, in the
spirit of the Hubbard-Stratonovich transformation \cite{Fadeev}, the
fermionic lagrangian above is turned into the `fermion-boson' model.

\begin{equation}
\label{eq2} {\cal L}=\sum_\sigma \Psi _\sigma ^{+}\left( i\partial _t\text{ }%
+ \text{ }\frac 1{2m^{*}}\nabla ^2\text{ }+\text{ }\mu _\sigma \right) \Psi
_\sigma \text{ }+\frac 12\phi ^2\text{ }-\text{ }\sqrt{U}\sum_\sigma \Psi
_\sigma ^{+}\Psi _\sigma \phi 
\end{equation}
with a Yukawa-type interaction which is bilinear in the fermion fields.

With this simplified model we can now proceed performing perturbation theory
expansions for the full fermion Green's function $G^\sigma $ and the boson
`propagator' $D^{\sigma \sigma ^{\prime }}$. The diagrams for the $G^\sigma $
and $D^{\sigma \sigma ^{\prime }}$ are depicted up to $3rd$ loop order in
Figures 1 and 2. Note that they are strongly spin dependent. Using their
corresponding Schwinger-Dyson equations they can be formally given in the
closed form:

\begin{equation}
\label{eq3}G^\sigma \left( {\bf p},p_0\right) =\left[ p_0-\left( \epsilon
\left( {\bf p}\right) -\mu _\sigma \right) -\Sigma _\sigma \left( {\bf p}%
,p_0\right) \right] ^{-1},
\end{equation}
where $\Sigma $ is the spin $\sigma $ irreducible self-energy and similarly

\begin{equation}
\label{eq4} D^{\sigma ,-\sigma }\left( {\bf p},p_0\right) =\left[ 1-U^2\Pi
_\sigma \left( {\bf p},p_0\right) \Pi _{-\sigma }\left( {\bf p},p_0\right)
\right] ^{-1}, 
\end{equation}

\begin{equation}
\label{eq4a}D^{\sigma ,\sigma }\left( {\bf p},p_0\right) =\frac{U\Pi
_{-\sigma }\left( {\bf p},p_0\right) }{1-U^2\Pi _\sigma \left( {\bf p}%
,p_0\right) \Pi _{-\sigma }\left( {\bf p},p_0\right) },
\end{equation}
where $\Pi _{\sigma \left( -\sigma \right) }$ is the proper polarization
function produced by the $\sigma \left( -\sigma \right) $ spin fermions. The
exact $\Sigma $ and $\Pi $ are shown diagrammatically in Figure 3.

Let us consider the polarization function in more detail. If we neglect the
vertex corrections and the self-energy insertions, the non-interacting
polarization function reduces to

\begin{equation}
\label{eq5}-i\Pi _\sigma ^f\left( {\bf p},p_0\right)
=-\int\limits_qiG_f^\sigma \left( {\bf q+p},q_0+p_0\right) iG_f^\sigma
\left( {\bf q},q_0\right) 
\end{equation}
where the free fermion Green's function $G_f^\sigma $ is given by

\begin{equation}
\label{eq6} G_f^\sigma \left( {\bf q},q_0\right) =\frac 1{q_0-\left(
\epsilon \left( {\bf q}\right) -\mu _\sigma \right) +i\delta sgn\left(
\epsilon \left( {\bf q}\right) -\mu _\sigma \right) } 
\end{equation}
with $\int_q=\int \frac{dq_0}{2\pi }\int \frac{d^2q}{\left( 2\pi \right) ^2}$
in $2+1$ dimensions.

Performing the contour integration in eqn.(\ref{eq5}) , we obtain that

\begin{equation}
\label{eq7} Re\Pi _\sigma ^f\left( {\bf p},p_0\right) ={\cal P}\int\limits_{%
{\bf q}}\Theta \left( \sqrt{2\mu _\sigma }-q\right) \left[ \frac
1{p_0+\omega _{{\bf qp}}}-\frac 1{p_0-\omega _{{\bf qp}}}\right] , 
\end{equation}
\begin{equation}
\label{eq7a} 
\begin{array}{c}
Im\Pi _\sigma ^f\left( 
{\bf p},p_0\right) =\pi \int\limits_{{\bf q}}\Theta \left( \left| {\bf p+q}%
\right| -\sqrt{2\mu _\sigma }\right) \Theta \left( \sqrt{2\mu _\sigma }%
-q\right) \times \\ \left[ \delta \left( p_0+\omega _{{\bf qp}}\right)
+\delta \left( p_0-\omega _{{\bf qp}}\right) \right] , 
\end{array}
\end{equation}
where $\omega _{{\bf qp}}=\epsilon \left( {\bf p}+{\bf q}\right) -\epsilon
\left( {\bf q}\right) $, being the single-particle energy difference and $%
\int_q=\int \frac{d^2q}{\left( 2\pi \right) ^2}$.

Let us consider a quasiparticle excitation defined around the same point on
the Fermi surface. For convenience we choose this point to lie along the
$y$-direction \cite{Wilczek}. In the vicinity of this point the single particle
energies becomes

\begin{equation}
\label{eq8}\epsilon \left( {\bf q}\right) =\frac 1{2m^{*}}\left(
q_x^2+q_y^2\right) \cong \frac 1{2m^{*}}q_x^2+\mu _\sigma +v_F^\sigma \left(
q_y-k_F^\sigma \right) 
\end{equation}
where $v_F^\sigma $ is the Fermi velocity for the $\sigma $-spin particle and

\begin{equation}
\label{eq9}\left\{ 
\begin{array}{c}
k_F^\sigma -\lambda \leq q_y\leq k_F^\sigma +\lambda  \\  
\\ 
-\sqrt{2k_F^\sigma \lambda }\leq q_x\leq \sqrt{2k_F^\sigma \lambda }
\end{array}
\right. 
\end{equation}
with $\lambda /k_F^\sigma \ll 1$.

As is well known from renormalization group analysis the relevant
interaction for quasiparticle arise when they are near the same point in the
Fermi surface. This means that for the quasiparticle defined around the
selected point along the $y$-direction the quasiparticle states that matter
are those located in the neighborhood regime defined according to eqn.(\ref
{eq9}). Thus if $G_f^\sigma $ refers to the single-particle excitation

\begin{equation}
\label{eq10}G_f^\sigma \left( {\bf q},q_0\right) \cong \frac 1{q_0-\frac
1{2m_\sigma ^{*}}q_x^2-v_F^\sigma \left( q_y-k_F^\sigma \right) \pm i\delta
},
\end{equation}
the single particle states it relates to are situated in vicinity of the
point $\left( 0,\sqrt{2m_\sigma ^{*}\mu _\sigma }\right) $ and we can
therefore make the approximation

\begin{equation}
\label{11}\sum_{k_F^\sigma -\lambda \leq \left| {\bf q}\right| \leq
k_F^\sigma +\lambda } 1 = 4\pi k_F^\sigma \lambda = \pi\sqrt{\frac{%
k_F^\sigma }{2\lambda }}
\sum_{ \stackrel{\scriptstyle 
 -\sqrt{2k_F^\sigma \lambda }\leq q_x\leq \sqrt{2k_F^\sigma \lambda }
 }
 {k_F^\sigma -\lambda \leq q_y\leq k_F^\sigma+\lambda}
} 1.
\end{equation}

Using this approximation to polarization function produced by the relevant
single-particle excitations reduces to

$$
Re\Pi _\sigma ^f\left( {\bf p};p_0\right) =\pi \sqrt{\frac{k_F^\sigma 
}{2\lambda }}\int\limits_{-\sqrt{2k_F^\sigma \lambda }}^{\sqrt{2k_F^\sigma
\lambda }}\frac{dq_x}{2\pi }\int\limits_{k_F^\sigma -\lambda }^{k_F^\sigma
}\frac{dq_y}{2\pi } 
$$

\begin{equation}
\label{12}\left[ \frac 1{p_0+\frac 1{2m_\sigma ^{*}}p^2+\frac{v_F^\sigma p}{
\sqrt{2}}+\frac{pq_x}{\sqrt{2}m_\sigma ^{*}}}-\frac 1{p_0-\frac 1{2m_\sigma
^{*}}p^2-\frac{v_F^\sigma p}{\sqrt{2}}-\frac{pq_x}{\sqrt{2}m_\sigma ^{*}}%
}\right] 
\end{equation}
where we take ${\bf p}=\frac 1{\sqrt{2}}\left( p,p\right) .$

Evaluating this integral we find for $p=2\sqrt{k_F^\sigma \lambda }$ and
small values of $p_0$

\begin{equation}
\label{eq13} Re\Pi _\sigma ^f\left( p=2\sqrt{k_F^\sigma \lambda };p_0\right)
= \frac{m^{*}}{\pi }\ln \left( \frac{4k_F\lambda }{m^{*}p_0}\right) , 
\end{equation}

This logarithmic singularity for $p_0\rightarrow 0$ requires the
redefinition of all fields and coupling parameters involved in the model.
Using renormalization theory one way to go forward is to introduce bare
fields and bare coupling such that

\begin{equation}
\label{eq14} 
\begin{array}{c}
\phi _0=Z_3^{1/2}\phi \\ 
\Psi _0=Z_2^{1/2}\Psi \\ 
\sqrt{U_0}= 
\frac{Z_1\sqrt{U}}{Z_2Z_3^{1/2}}=Z_3^{-1/2}\sqrt{U} \\ \mu _0=\mu -\triangle
\mu , 
\end{array}
\end{equation}
using the Ward identity result \cite{Mahan} $Z_1=Z_2$.

All the infinities which arise in perturbation theory are appropriately
included in the bare quantities, in a way that, if we start with the
corresponding bare lagrangian ${\cal L}_0$ instead of the one given by eqn. (%
\ref{eq2}) the resulting renormalized propagators $G_R$, $D_R$ , the vertex
function $\Gamma _R$ and the renormalized coupling $U$ are finite.

Let us consider the full bare boson `propagator' $D_0^{\sigma ,\sigma }$.
Using the Schwinger-Dyson result from eqn. (\ref{eq4a}) we have that

\begin{equation}
\label{eq15}D^{\sigma ,-\sigma }_0\left( {\bf p},p_0\right) =\frac
1{1-U_0^2P_{\sigma ,-\sigma }\left( {\bf p},p_0\right) }=Z_3D_R^{\sigma
,-\sigma }\left( {\bf p},p_0\right) ,
\end{equation}
with $P_{\sigma ,-\sigma }\left( p=2\sqrt{k_F^\sigma \lambda },p_0\right) =
\frac{m_0^{*2}}{\pi^2}\ln {}^2\left( \frac{4k_F\lambda }{m^{*}p_0}\right) $
, where for simplicity, we take $k_F^\sigma =k_F^{-\sigma }=k_F$.

If we define the renormalized `propagator' $D_R$ in a way that $%
D_R\rightarrow 1$, at $p=2\sqrt{k_F^0\lambda }$, when $p_0\rightarrow \omega 
$ , some characteristic energy scale, we obtain

\begin{equation}
\label{eq16} D^{\sigma ,-\sigma }_0\left( p=2\sqrt{k_F\lambda };p_0\right) = 
\frac{Z_3}{1-\frac{m_0^{*2}}{\pi^2}U_0^2\ln \left( \frac{\left(
4k_F\lambda /m_0^{*}\right) ^2}{\omega p_0}\right) \ln \left( \frac \omega
{p_0}\right) Z_3}, 
\end{equation}
where 
\begin{equation}
\label{eq17} Z_3^{-1}=1-c_0U_0^2\ln {}^2\left( \frac{4k_F\lambda /m_0^{*}}%
\omega \right) , 
\end{equation}
with $c_0=m_0^{*2}/\pi^2>0.$

Using this result the bare and the renormalized coupling constants can be
readily related to each other to give

\begin{equation}
\label{eq18} \frac{U_0}{1-c_0U_0^2\ln {}^2\left( \frac{4k_F\lambda /m^{*}}%
\omega \right) }=U 
\end{equation}
or, equivalently

\begin{equation}
\label{eq19} U_0=U\left( 1-cU^2\ln {}^2\left( \frac{4k_F\lambda /m^{*}}%
\omega \right) +O\left( U^3\right) +...\right) , 
\end{equation}
with $c$ given in terms of the corresponding renormalized effective mass $%
m^{*}$ and renormalized $k_F$. Here $\omega $ is an arbitrary scale
parameter. However since the bare coupling is independent of any variation
in this scale we must have $\omega \partial U_0/\partial \omega =0$. From
this we obtain the renormalization group equation for the renormalized
coupling $U$:

\begin{equation}
\label{eq20} \omega \frac{\partial U}{\partial \omega }=-2c U^3\ln \left( 
\frac{4k_F\lambda /m^{*}}\omega \right) +... 
\end{equation}

Neglecting the higher order terms, this equation can be easily integrated to
give

\begin{equation}
\label{eq21}U^2\left( \omega \right) =\frac{U^2\left( \Omega \right) }{%
1-2c U^2\left( \Omega \right) \ln \left( \frac{\left( 4k_F\lambda
/m^{*}\right) ^2}{\Omega \omega }\right) \ln \left( \frac \Omega \omega
\right) }
\end{equation}
where $\Omega $ is some upper energy value. If we take for simplicity $%
\Omega =4k_F\lambda /m^{*}$ this reduces simply to

\begin{equation}
\label{eq22}U^2\left( \omega \right) =\frac{U^2\left( \Omega \right) }{%
1-2c U^2\left( \Omega \right) \ln {}^2\left( \frac \Omega \omega \right) }
\end{equation}

Even if $U\left( \Omega \right) $ is sufficiently small, $U\left( \omega
\right) $ grows stronger for low values of $\omega $ and we are driven
outside the domain of the validity of perturbation theory when $%
2c U^2\left( \Omega \right) \ln {}^2\left( \frac \Omega \omega \right)
\rightarrow 1.$ This takes place, for this case at an energy scale $\omega = 
\frac{4k_F\lambda }{m^{*}}\exp \left( -\frac{\pi}{\sqrt{2}m^{*}U}\right) $.
Consequently the Fermi liquid infrared fixed point is physically
unattainable in this regime since the renormalized coupling becomes
infinitely large and negative as $\omega \rightarrow 0$.

In conclusion we show that the free polarization function in $d=2$ spatial
dimensions is logarithmic divergent for quasiparticles which are
near the same
point on the Fermi surface. This singularity can be regularized if we apply
renormalization theory. We do this defining bare fields and bare coupling
constants which absorb all the divergences in all orders of perturbation
theory. In particular, using the corresponding Schwinger-Dyson equation, we
calculate the bare boson `propagator' $D^{\sigma ,-\sigma }_0$ which measures
essentially the effective interaction between the $\sigma $ and $-\sigma $
quasiparticles. From it we establish how the bare and the renormalized
couplings $U_0$ and $U$ relate to each other. Taking into account that the
bare coupling must be independent of the variations of any scale parameter
introduced in the theory we then proceed by constructing the renormalization
group equation for the renormalized coupling constant. If we neglect higher
order contributions the RG equation can be easily integrated. As a result of
this we show that even if the coupling constant we start with is
sufficiently small, the coupling grows stronger and stronger as the scale
parameter approaches zero. This behavior drives the theory outside the
domain of validity of perturbation theory and invalidates Fermi liquid
theory since its trivial infrared fixed point is unapproachable in this
regime.

Although the model under considerations is rather simple,
it can be related to the Hubbard model in the long wavelength limit.
As is well know for the $d=2$ Hubbard model the fermion spectral
function becomes significantly
broader as a result of the increase in the Hubbard $U$ \cite
{Preuss,Matsu,Saik}. Consequently the Fermi surface is no longer sharply
defined. However the relevant interactions for quasiparticle take place when
they are near the same point on the Fermi surface. Defining the relevant
region in the thin shell around the non-interacting Fermi surface for a
quasiparticle defined near a given point in $S_F$ we calculate the
polarization function and show that it
can be singular for low energy value. Using
renormalization theory we regularize our model defining new renormalized
coupling and fields. New features are now present. Contrary to what happens
in Fermi liquid theory the renormalized coupling constant grows indefinitely
for low values of the corresponding energy scale parameter. Thus there is no
trivial infrared fixed point and the system is driven to a non-perturbative
regime when it approaches the Fermi surface point. There remains to be
discussed the physical nature of this non-Fermi liquid state\cite{Varma2}
and the formation of bound states of quasiparticles which may lead to the
onset of superconductivity. These problems are presently under consideration
and will be addressed elsewhere.

\begin{figure}
\caption{
Fermion single-particle Green's function $G_{\sigma}$.
}
\label{Fig1}
\end{figure}

\begin{figure}
\caption{
Auxiliary boson full-`propagators' $D_{\sigma,-\sigma}$
and  $D_{\sigma,\sigma}$.
}
\label{Fig2}
\end{figure}

\begin{figure}
\caption{
Exact fermion irreducible self-energy $\Sigma_{\sigma}$
and proper polarization function $\Pi_{\sigma}$ given in terms of
vertex function $\Gamma$ and the irreducible
particle-hole scattering amplitude $K_{\sigma,-\sigma}$.
}
\label{Fig3}
\end{figure}

\end{document}